**High electrical conductivity of single metal-organic chains**

Pablo Ares, Pilar Amo-Ochoa, José M. Soler, Juan José Palacios, Julio Gómez-Herrero* and Félix Zamora*

Dr. P.A., Prof. J.M.S., Dr. J.J.P., Prof. J.G.H.
Departamento de Física de la Materia Condensada
Universidad Autónoma de Madrid, Madrid E-28049 (Spain)
E-mail: julio.gomez@uam.es

Dr. P.A.O., Dr. F.Z.
Departamento de Química Inorgánica
Universidad Autónoma de Madrid, Madrid E-28049 (Spain)
E-mail: felix.zamora@uam.es

Prof. J.M.S., Dr. J.J.P., Prof. J. G. H., Dr. F.Z.
Condensed Matter Physics Center (IFIMAC)
Universidad Autónoma de Madrid, Madrid E-28049 (Spain)

Dr. F.Z.
Instituto Madrileño de Estudios Avanzados en Nanociencia (IMDEA-Nanociencia)
Cantoblanco, Madrid E-28049 (Spain)

**Abstract**

Molecular wires are essential components for future nanoscale electronics. However, the preparation of individual long conductive molecules is still a challenge. MMX metal-organic polymers are quasi-one-dimensional sequences of single halide atoms (X) bridging subunits with two metal ions (MM) connected by organic ligands. They are excellent electrical conductors as bulk macroscopic crystals and as nanoribbons. However, according to theoretical calculations, the electrical conductance found in the experiments should be even higher. Here we demonstrate a novel and simple drop-casting procedure to isolate bundles of few to single MMX chains. Furthermore, we report an exponential dependence of the electrical resistance of one or two MMX chains as a function of their length that does not agree with predictions based on their theoretical band structure. We attribute this dependence to strong Anderson localization originated by structural defects. Theoretical modeling confirms that the current is limited by structural defects, mainly vacancies of iodine atoms, through which the current is constrained to flow. Nevertheless, measurable electrical transport along distances beyond 250 nm surpasses that of all other molecular wires reported so far. This work places in perspective the role of defects in one-dimensional wires and their importance for molecular electronics.

Circuit miniaturization is approaching sizes where quantum uncertainties will make transistors unreliable.[1] Actually, current state-of-the-art silicon technology enables up to 100 million transistors/mm$^2$ being the typical size for a transistor about 100 nm, although it only allows for the fabrication of relatively simple circuits. In contrast, molecular electronics offers a prospect of size reduction by using molecular building blocks to fabricate electronic components. Molecules like DNA[2] and DNA derivatives, including origami[3] and metallized DNA,[4] exhibit self-assembling capabilities that suggest a potential for a substantial increase in the complexity of molecular circuitry. Therefore, beyond miniaturization, the fabrication of very complex architectures is of high interest in molecular electronics.[5] Many organic molecules and polymers have already been tested as molecular wires, but very few have shown a reasonably low electrical resistance.[6] As a general rule,[7] they show a high resistance that increases exponentially with the molecular length.[6b] For short organic molecules of 1-3 nm, this behaviour simply originates from an energy level alignment with the chemical potential in an HOMO-LUMO gap.[8] For longer molecules, other mechanisms such as electron hopping come into play.[6b] Thus, conventional metallic conductors in the diffusive regime (not coherent), such as gold nanowires, exhibit a weak *linear* dependence of the resistance *vs.* length (Supporting Information for additional details). Typical organic polymer conductors show very low intrinsic conductivity. This increases using chemical dopants in bulk[9] but not in isolated chains, like those addressed in this work. Another possible mechanism for a high resistance can be understood within the theory of Anderson localization. According to this theory, a one-dimensional conductor, whose carrier wave functions preserve coherence over sufficiently long distances, becomes an insulator when even a low concentration of random defects is present.[10] The resistance thus increases *exponentially* with the length of the conductor divided by a characteristic distance related to the localization length. Exceptions to short localization lengths are a recent work by Slinker *et al.* reporting transport in self-assembled DNA layers with a localization length in the range of 30 nm,[11] and few more works on metal-organic molecules[12] and polymers, the former by Tuccitto *et al.*,[13] based on a layer by layer assembly of metal ions with terpyridine ligands, and the latter by us on MMX nanoribbons.[14] Therefore, the combination of metal ions with organic molecules, despite its limited development, has already shown promising results.[15] Importantly, density functional theory (DFT) calculations predicted that some MMX polymers (in particular the one used in this work) present an electronic band structure characteristic of a metal and hence one would expect a linear dependence of the electrical resistance *vs.* length.

MMX chains are coordination polymers based on the assembly of two dimetallic entities, *e.g.* [$Pt_2(dta)_4$] and [$Pt_2(dta)_4I_2$] (*dta* = ditiocarboxylate), some of them showing interesting electronic properties in bulk[16] (**Fig. 1**). The relative weakness of the coordination bonds allows the self-assembly of these polymers,[17] but it also facilitates the presence of defects like iodine vacancies. The first attempts to synthesize nanometric forms of these polymers involved a non-conventional and tedious procedure, based on sublimation from MMX crystals. This is not very appropriate for molecular electronics[17] because it leads to nanoribbons of hundreds of individual MMX chains.[14] Their electrical transport over distances beyond ~ 200 nm is highly affected by the presence of structural defects and it is dominated by interchain electron transfer, leading to diffusive transport.[14]

Recently, we have shown that crystals of [Pt$_2$(*dta*)$_4$I]$_n$ can dissociate into a solution of [Pt$_2$(*dta*)$_4$] and [Pt$_2$(*dta*)$_4$I$_2$] that, under suitable conditions, reverse to the polymeric [Pt$_2$(*dta*)$_4$I]$_n$ chains.[18] Here we report the isolation of single chains of [Pt$_2$(EtCS$_2$)$_4$I]$_n$ (Et = CH$_2$-CH$_3$) on surfaces, by simple drop-casting, and their subsequent electrical characterization. A dichloromethane solution of monocrystals of [Pt$_2$(EtCS$_2$)$_4$I]$_n$ was deposited by drop-casting on SiO$_2$ substrates at room temperature and under atmospheric conditions. Inspection by optical microscopy (**Fig. 2a**) reveals the formation of long (over one micron) fibre-like aggregates, as well as lower concentrations of smaller aggregates. There is a clear correspondence in optical images between colour, size, and concentration (brown>orange>yellow>blue for larger/denser to smaller/sparser). Atomic Force Microscopy (AFM) in regions with low density (blue fibres in the optical image) shows that these nanostructures have lengths of tens of microns and heights of ~ 20-100 nm (Fig. 2b), in the same range reported for MMX nanoribbons obtained by crystal sublimation.[14] Further inspection shows that the nanostructures at the lowest density regions exhibit lengths of 100-800 nm and heights of 1.7-10 nm (Fig. 2c,d). Since the distance between individual chains in MMX crystals is about 1 nm (Fig. 1), we ascribe these nanostructures to bundles of very few MMX chains. Scanning Electron Microscopy (SEM) images (Figure S1 in the Supporting Information) show excellent agreement with AFM data, but do not allow resolution of the chains down to the single-molecule level.

**Fig. 3** shows electrical transport measurements on single chains of [Pt$_2$(EtCS$_2$)$_4$I]$_n$ protruding from a gold electrode. To this end, a macroscopic gold electrode was evaporated on a previously prepared sample, resulting in a number of fibres partially covered by this fixed electrode (Fig. 3a). A conductive AFM (C-AFM) tip is used as a second mobile electrode to measure the current through the chains as a function of their length. The experimental procedure is as follows: (i) a non-invasive topography image is acquired to locate the molecules, (ii) the metallized AFM tip is brought into mechanical and electrical contact with the MMX chains; (iii) a bias ramp is applied to the tip while the current is monitored; and (iv) the tip is withdrawn and moved to the next point of the chain (see Methods for details). During I-V acquisition we have thoroughly checked that no current is measured (within experimental noise) on the silicon oxide substrate in the vicinity of the chains. Fig. 3b is an AFM topographic image of one/two of the $[Pt_2(EtCS_2)_4I]_n$ chains protruding from the gold contact. The inset shows a profile across the final part of the chains, showing a height of ☐☐1 nm, compatible with a single molecule chain (see section "Geometrical Considerations" in the Supporting Information for a more detailed discussion of this issue). Therefore, we estimate that this final part is made of at most two single chains, thus allowing single-molecule electrical measurements.

Fig. 3c displays the electrical resistance $R$ as a function of the distance $L$ between electrodes, obtained for the chains in Fig. 3b. At first glance we observe a significant current at distances as long as 280 nm for a single molecule experiment. Nevertheless, even discounting the tip-sample contact resistance (see Supporting Information), the measured resistance is still much higher than the quantum resistance $R_0$ predicted by DFT calculations for this system. They show a metallic band structure[19] with a spin-degenerate half-filled band crossing the Fermi level along the ☐-A direction, the chain axis. This is not surprising within the single-particle DFT description, given the odd number of electrons in the monomeric unit. According to Landauer's theory, a one-dimensional band that crosses the Fermi level should result in a quantum of resistance $R_0 = h/(2e^2) \approx 13$ kΩ. This value refers to ballistic charge transport and does not take into account other effects such as the electrode-chain contact resistance, electron-phonon interactions, scattering with the substrate, or the presence of structural defects along the chains. Our results point towards the presence of some of these effects increasing the electrical resistance along a single MMX chain. In addition, there is a clear non-linear dependence that suggests electron-defect interactions as the main source of resistance in these chains.

An exponential $R(L)$ dependence, observed in carbon nanotubes,[20] was attributed to Anderson localization caused by the interaction of electrons with structural defects. Similarly, we fit our data to an exponential expression (equation 1)

$$R = R_c + \frac{R_0}{N_0}\exp(L/L_0)$$

(1)

where we assume that the chains are electrically isolated. In this expression $R_c$ accounts for the electrode-chain contact resistance, $R_0$ is the above mentioned quantum resistance, $N_0$ is the number of chains contacted by the AFM tip, and $L_0$ is the localization length. In Fig. 3b, the AFM topography shows that $N_0$ can be 1 or 2 (Figs. S2 and S3). Numerical fitting of the data shown in Fig. 3c (see Supporting Information Table S1 and Fig. S4 for more details) to equation (1), with $N_0 = 2$ and a bias voltage of 0.5 V, leads to $R_c \approx 800$ MΩ and $L_0 \approx 18$ nm. According to Kelley et al.[21] the nonlinearity in the I-V curves (inset in Fig. 3c) can be explained by charge injection barriers at the tip-MMX junction, that make $R_c$ depend on $V$. In order to check the robustness of our fit, we have repeated it for higher voltages, finding that $L_0$ varies between ~ 17 and 23 nm (Fig. 3d). This variation is quite small, given the ranges of voltages and number of molecules, and it shows that our determination of the localization length is robust. To further assess the reliability of our measurements, Fig. S4a shows ~ 100 resistance values obtained from I-V curves in fibres composed of several chains (fibre diameters of 8-14 nm). The dispersion is quite moderate for this kind of nanoscale measurements. The *R*(*L*) dependence is consistent not only within the same fibre, but also among different fibres and AFM tips. Figure S4b displays *R*(*L*) curves for three representative fibres of similar geometries but different alkyl groups (*dta*). The increase in resistivity with the length of the alkyl group is consistent with the proposed conduction mechanism in sublimated molecules.[14a] In these, the current within individual chains is hampered by structural defects, but in fibres composed of few chains, electrons can hop to adjacent chains and continue moving. With longer alkyl groups, the hopping distance increases, and so it does the resistance.

A remarkable feature of these $Pt_2$-I-$Pt_2$ chains is that the spatially-averaged current density across the whole chain cross-section (using a chain radius of ~ 1 nm) is ~ $10^5$ A $cm^{-2}$ (see Fig. S6 for details). This is ~ $10^{10}$ times larger than that reported by Tuccitto *et al.*[13] for single molecules of a metal-based terpyridine polymer. This significant difference in current densities suggests a much lower contact resistance in our experiments. Furthermore, our current density values come from measurements with a distance between electrodes of 80 nm, while those in reference [13] correspond to lengths below 15 nm.

In molecular electronics, the *β*-decay factor of coherent electrical transport across short molecules is commonly used as an indicator of the charge motion efficiency (I~ $e^{-\beta L}$). In insulating and semiconducting polymers (the vast majority of them), beta decays arise from coherent dispersion by a periodic potential. In our intrinsically metallic system, the current decay is a consequence of Anderson localization, produced by coherent reflection and transmission from random defects. Although the two mechanisms are different, both result in an exponential decay of the electrical current with length and for the sake of comparison we will characterize them by a *localization* length $\beta = 1/L_0$. Thus, the lower (higher) the $\beta$ ($L_0$) value, the *better* the electrical transport.

Although this decay factor and its associated localization length do not have the same physical origin in short molecules and in the Anderson model of long defective chains, we will use it indistinctly. In our case $\beta \sim 5 \cdot 10^{-3}$ Å$^{-1}$ while references [13] and [14a] report slightly lower values for larger systems, whose synthesis usually has a lower probability for creation of defects. Barton *et al.*[11] also reported conductance over 34 nm in DNA layers using indirect evidence from electron transfer rates in solution (they do not report I-V characteristics). However, as far as we know, DFT calculations in double stranded DNA show an insulating band structure and hence the electrical conduction should have a different mechanism than in MMX polymers.[14a] Table S2 summarizes the comparisons.

To support our conjecture that Anderson localization, induced by structural defects, is the main mechanism behind the observed exponential dependence of resistance with length (Eq. (1)), we have carried out quantum transport calculations at two levels of sophistication. First we have computed the quantum conductance for a single defect in an otherwise perfect MMX wire. We used the ANT.1D code[22] and the Green's function formalism for quantum transport,[23] with DFT matrix elements computed with SIESTA,[24] (see Methods for details). We have considered three possible defects: 1) a kink; 2) an OH molecule substituting an iodine atom; and 3) an iodine vacancy (**Fig. 4a**). The kink accounts for the effects of substrate roughness. The other defects may arise in MMX synthesis from solution.[18] Thus, as shown in reference [25], the assembly of [Pt$_2$(EtCS$_2$)$_4$] building blocks may lead to iodine vacancies. Equally, we consider the possible substitution of some iodide anions by hydroxyl groups. The calculated electron transmission across these three defects, as a function of electron energy, is shown in Fig. 4b. The kink does not appreciably change the transmission within ± 0.5 eV of the Fermi energy $E_F$. The OH substitution decreases the transmission at $E_F$ by a factor 0.9. Finally, the vacancy has the largest effect, reducing the transmission by a factor 0.7. Therefore, we will focus on this last defect.

The conductance for long MMX wires, with many vacancy defects, has also been calculated with ANT.1D, but now using a single-orbital tight-binding model, with parameters fitted to reproduce the single-defect transmission. This is justified because the only conductance channel in the MMX wire is made of a single band. In Fig. 4c,d we show results for two concentrations of defects, placed at random positions and for segments of increasing length along three different wires. We do not have averaged the conductance over different positions of the defects to mimic the actual experiment with the tip moving along the same wire. As expected in phase-coherent calculations, and ignoring the inherent fluctuations, the conductance follows an exponential decay with wire length. According to these calculations, a percentage of vacancies in the range 6-8% would be responsible for the observed decay length. Considering our previous results on MMX nanoribbons, this concentration of defects is perfectly compatible with those observed by direct sublimation of MMX microcrystals.[14]

One of the most important conclusions of this work is that coordination polymers, also known as metal-organic polymers, are prone to defects. As one-dimensional conductors, they are very sensitive even to a low density of defects, hence they will be always in the strong Anderson localization regime for long enough lengths. Arguably, the localization length can be of the order of tens of nm (in our case 20 nm), still much longer than the electrical current decay with length observed in the vast majority of organic molecules and polymers. We present three different experimental facts to support this conclusion: *i*) In ref. [14] we observed the influence of defects in nanoribbons prepared by sublimation. In this work, we also observe the influence of defects but now at the single molecule level and for MMX prepared by drop casting, a very different approach. *ii)* The β-decay reported in ref. [13] for terpyridine is ~1/30 nm$^{-1}$, similar to our 1/$L_0$. Therefore, the density and strength of the defects should be similar to that presented in our MMX nanoribbons and individual fibres. *iii*) The electrical conductance measured in macroscopic MMX crystals is slightly higher than what is measured at the nanoscale,[14a] therefore suggesting a slightly lower density of defects. As macroscopic crystals are grown close to thermodynamic equilibrium one should expect a lower density of defects than in nanoscopic MMX structures where the growth is dominated by kinetic (fast growth).

The unavoidable presence of defects in coordination polymers can be understood in terms of the moderated strength of coordination bonds. Self-assembly always implies weak bonds but the counter part is the presence of defects. The situation for carbon nanotubes, a paradigmatic example, is completely different: they exhibit strong covalent bonds with a very low density of defects resulting in a linear dependence of the resistance *vs.* length even for nanotubes longer than 10 μm. Precisely, these strong covalent bonds make functionalization of carbon nanotubes very difficult.

The implication of this work is that despite the influence of defects in the electrical conductance, coordination polymers are one of the best options for molecular wires at long distances as they exhibit a good balance between long-range transport and possible functionalization. Future routes to repair defects and functionalize this general class of polymers might be an important step forward for molecular electronics.

**Experimental**

*Synthesis of [Pt$_2$(EtCS$_2$)$_4$I]$_n$.* It was prepared according to reference [16].

*AFM sample preparation. i)* Si substrates, with a SiO$_2$ layer of 300 nm, were exposed to an oxygen plasma pre-treatment.[14a] *ii)* Crystals of [Pt$_2$(EtCS$_2$)$_4$I]$_n$ (0.2 mg) were dissolved in CH$_2$Cl$_2$ (2 mL) and then centrifuged (4000 rpm, 5 min.) at room temperature. *iii)* 30 µl were deposited on a substrate and left 40 s for adsorption. After this time, substrates were dried in an argon gas jet. Samples were first inspected using an optical microscope to find the areas where the concentration of fibres was adequate.

*Conductive Atomic Force Microscopy (C-AFM) characterization.* AFM measurements were carried out using a Cervantes Fullmode AFM from Nanotec Electronica SL. WSxM software (www. wsxm.es) was employed both for data acquisition and image processing.[26] Commercial Cr/Pt coated Budget Sensors cantilevers (www.budgetsensors.com), with nominal force constant of 3 N m$^{-1}$ and resonance frequencies of about 75 kHz, were used in the experiments. Nominal tip radius was 25 nm according to the data provided by the manufacturer (Budget Sensors: ElectriMulti75-G). Electrical characterization of the MMX fibres was performed by conductive AFM at room temperature in ambient conditions. To this end, a gold layer of 30 nm was evaporated, using a stencil mask, on previously prepared samples. This resulted in a number of fibres partially covered by this fixed electrode. The conducting AFM tip was used as a second mobile electrode to measure the current through the uncovered parts of the fibres, measuring the current *versus* voltage characteristics (I-V) as a function of the distance between the tip and the fixed gold electrode. From the I-V plots, the resistance *versus* length curve $R(L)$ was determined along the fibre. In order to avoid tip and sample damage, topographic images were always taken in dynamic mode.[21] The experimental setup enabled increasing the loading force applied by the mobile electrode to the nanoribbon, while I-V characteristics were measured, until an optimum contact resistance was reached. In this way, simultaneous force-distance $F(Z)$ and current-distance $I(Z)$ curves were obtained ($Z$ being the piezoelectric displacement along the perpendicular direction to the sample surface). The maximal current was measured for an applied force of ~ 70 nN. The same force was used for all the measurements reported in the manuscript.

*Density functional calculations.* The matrix elements for the transmission calculations were obtained with SIESTA, using the GGA-PBE functional and Troullier-Martins pseudopotentials in their fully non-local form. A basis of double-zeta valence orbitals plus single polarization orbitals (DZP) was used, with an energy shift parameter of 50 meV. Real-space integrals were performed with a mesh cutoff of 200 Ry. Only the gamma *k* point was used. Convergence tests were performed with larger and longer bases, and with larger mesh cut-offs. A single defect was included, of one of the three types explained previously, in the centre of supercells of 4, 6, and 8 unit cells, with the two extreme unit cells kept fixed at the non-defective geometries. The rest of the geometry, including the supercell length, was relaxed until all the atomic forces were smaller than 40 meV/Ang. For kink defects, the kink angle was constrained during the relaxation.

*Quantum transport calculations.* The phase-coherent conductance is given by the Landauer formula $G = (2e^2/h)T$, where

$$T = Tr[\Gamma_L G^a \Gamma_R G^r]$$

is the standard transmission. For a single defect, the matrix elements needed to build the advanced (*a*) and retarded (*r*) energy-dependent Green's function matrices (*G*) that appear in the trace operation (*Tr*) are obtained with SIESTA, as explained above. The energy-dependent matrices $\Gamma$, for coupling to the perfect semi-infinite left (*L*) and right (*R*) MMX wires on both sides of the defect, were also obtained from SIESTA calculations for perfect wires. We have discussed in the main text the results obtained for systems with 6 unit cells, but we have checked that the transmission functions obtained using 8 unit cells were nearly identical.

The same methodology was used in the single-orbital model calculations for wires with many defects. The model parameters (the on-site and nearest-neighbour hopping energies) were chosen to reproduce the position in energy and width of the conducting DFT-SIESTA band. Within this model, the defect sites representing the vacancies have a different on-site energy that reproduces the 0.7 transmission value at $E_F$ in a single defect calculation.

**Acknowledgements**
This work was supported by MINECO projects Consolider CSD2010-00024, MAT2016-77608-C3-1-P and 3-P, FIS2012-37549-C05-03, FIS2015-64886-C5-5-P, and FIS2016-80434-P. J.S., J.J.P., J. G. H. and F. Z. acknowledge financial support through The "María de Maeztu" Programme for Units of Excellence in R&D (MDM-2014-0377). The authors thank A. Gil for insightful discussions. J.J.P. also acknowledges the European Union structural funds and the Comunidad de Madrid under grant nos. S2013/MIT-3007 and S2013/MIT-2850; the Generalitat Valenciana under grant no. PROMETEO/2012/011, and the computer resources and assistance provided by the Centro de Computación Científica of the Universidad Autónoma de Madrid and the RES.

15

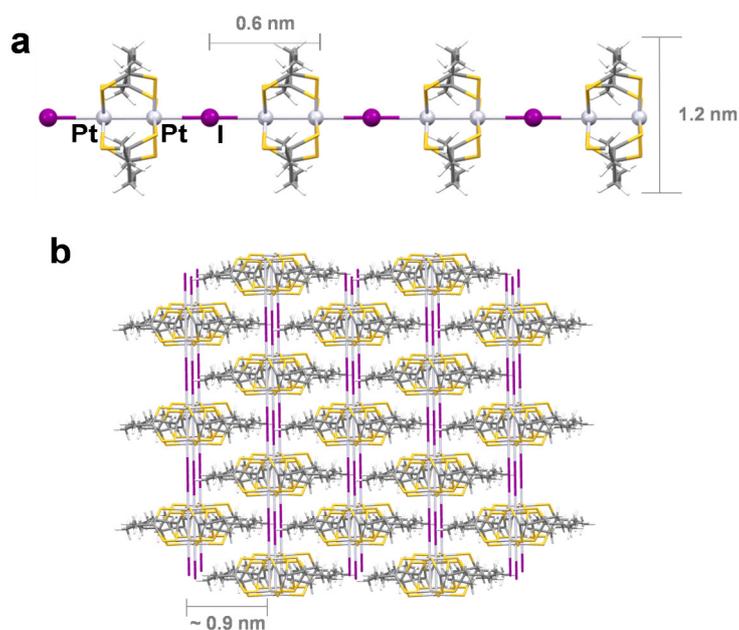

**Figure 1.** Crystal structure details of our MMX polymers. **a,** Schematic representation of a [$Pt_2(EtCS_2)_4I$]$_n$ (Et = $CH_2$-$CH_3$) chain. **b,** Crystal packing and selected distances of the same chains.

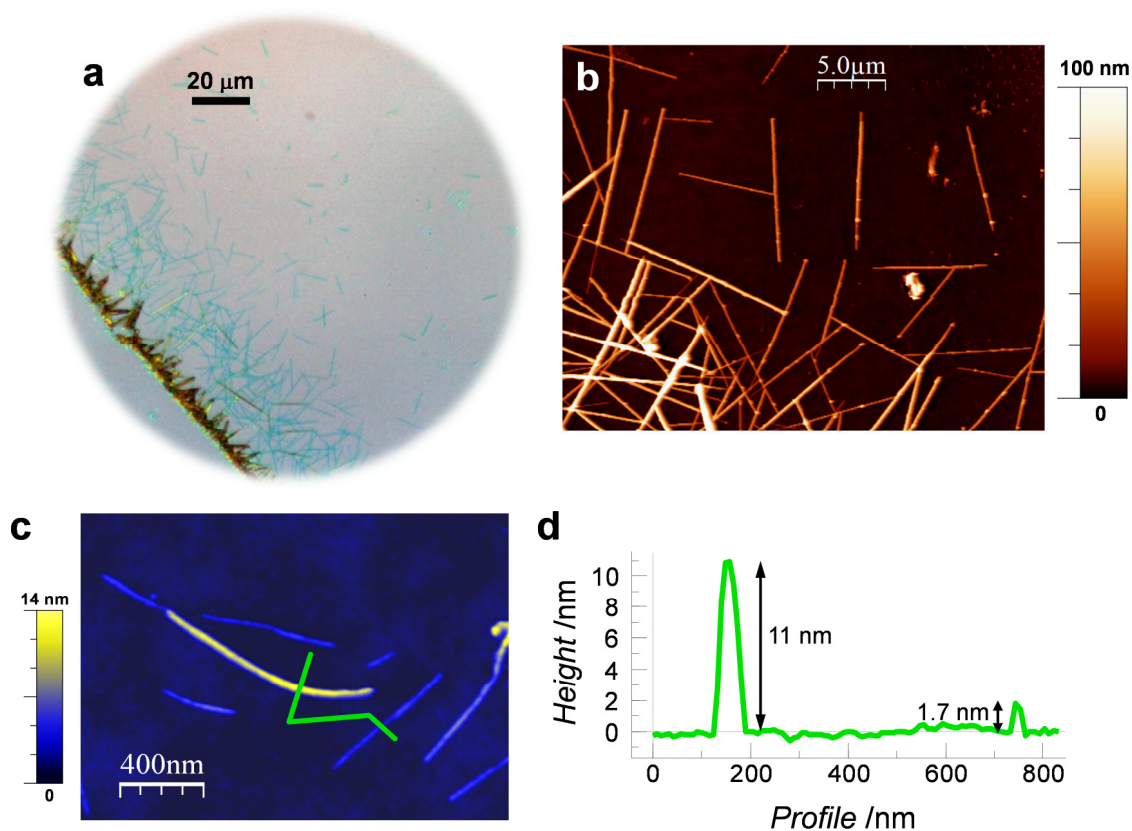

**Figure 2.** Morphological characterization of [$Pt_2(EtCS_2)_4I$]$_n$ fibres. **a,** Optical image of a drop-casted dichloromethane solution of the chains on a $SiO_2$ substrate. Thinnest fibres appear in blue. **b,** Topographic AFM image of the same substrate shown in **a,** in a region close to the border line of the large crystalline fibre-like aggregates. **c,** Topographic AFM image in a region far from the aggregates, where the fibres are almost not visible by optical microscopy. **d,** Height profile along the solid green line in **c**.

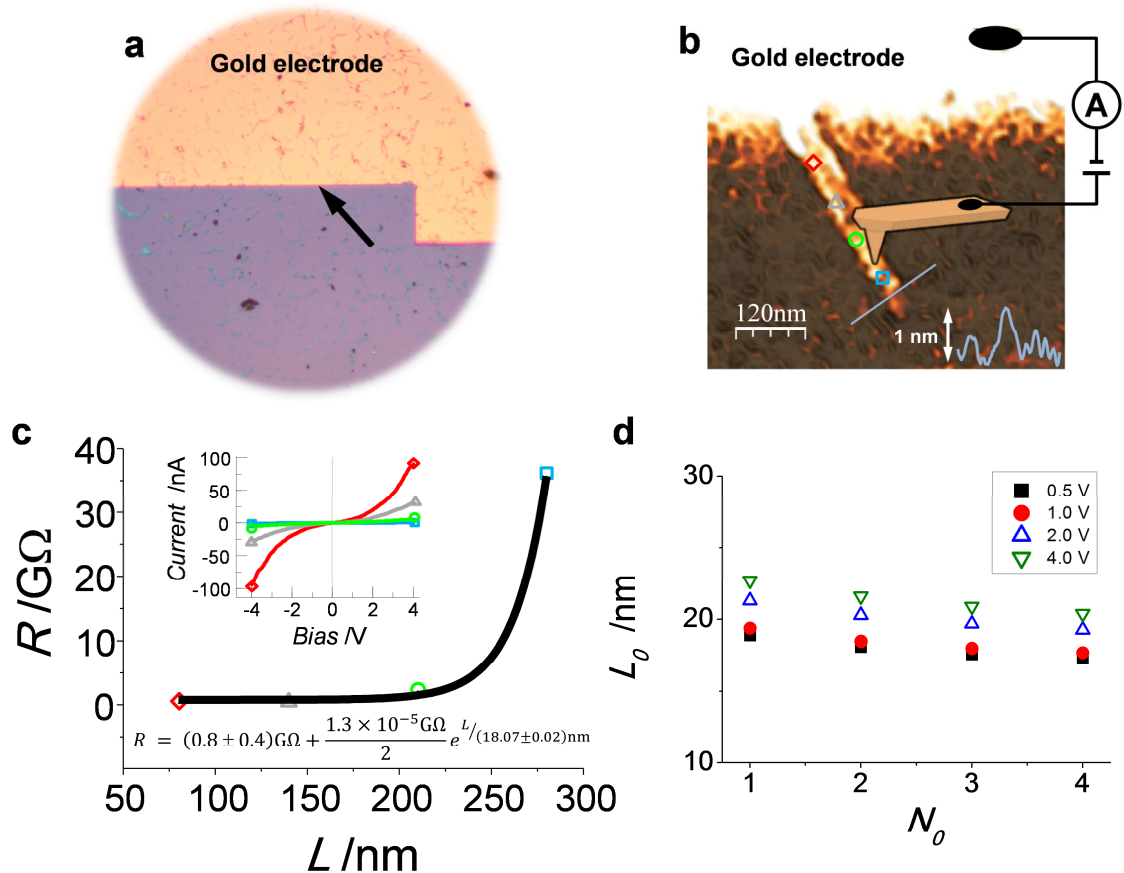

**Figure 3.** Single molecule electrical characterization of two [$Pt_2(EtCS_2)_4I$]$_n$ chains. **a,** Optical microscopy image of the gold electrode edge. The black arrow is pointing the region where the chains shown in **b** are located. **b,** Topographic AFM image of two single chains protruding from the gold electrode. Colour scale (from dark to bright) was adjusted to enhance the visibility of the chains. For clarity, the electrical circuit used in the AFM conductance experiments has been added to the AFM image. The inset shows the height profile along the solid blue line. The coloured symbols along the chains mark the spots of I-V measurements. **c,** Resistance *versus* length for the chains in **b**. The symbols represent resistances calculated from the I-Vs shown in the inset, at Bias = 0.5 V at the points shown in **b**. The black line is a fit to equation (1) considering two chains. **d,** Localization length for resistances obtained at 0.5, 1, 2 and 4 V Bias voltages as a function of the number of chains, $N_0$.

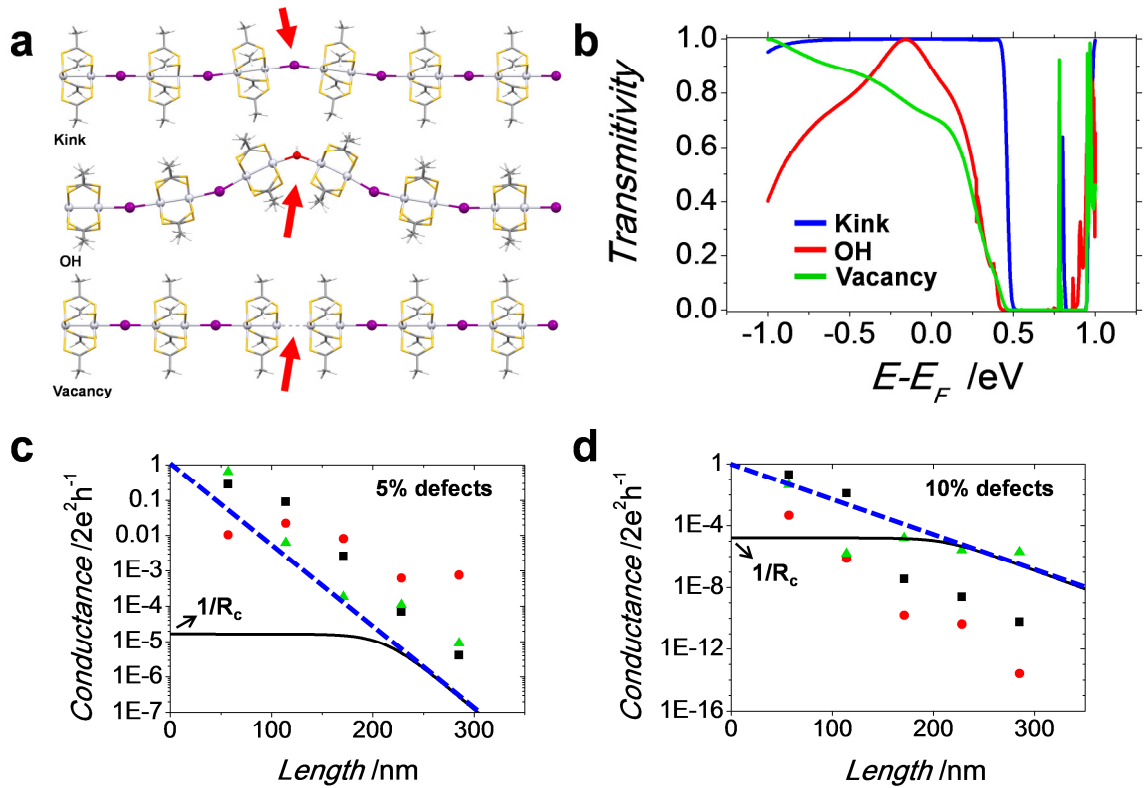

**Figure 4.** Theoretical calculations of $[Pt_2(MeCS_2)_4I]_n$ chains. **a,** Schematic relaxed structures of the three types of defects considered in our work (red arrows indicate the structural defect). From top to bottom: a kink, an OH group substituting an I atom, and an I vacancy. **b,** Transmission, as a function of energy, for a single defect of each of the three types shown in **a**. **c,** Conductance obtained for three different disorder realizations at a 5% concentration of vacancies (black squares, red circles and green triangles correspond to each of these realizations) in a single-orbital tight-binding model. The introduction of disorder mimics the random presence of vacancies along the chains. Large fluctuations can be appreciated on top of an overall exponential behaviour. The black line corresponds to the experimental result while the blue dashed line corresponds to an extrapolation of the measured conductance, assuming absence of experimental contact resistance. **d,** The same as in **c**, but for a larger concentration of vacancies (10%). For a 6-8% of vacancies we obtain a fairly good agreement between experimental data and calculations.

# Supporting Information

**MATERIALS AND METHODS**

*Scanning Electron Microscopy (SEM)*

Field emission scanning electron microscope Philips XL30 S-FEG. We have taken SEM images of different MMX fibres (**Figure S1**) showing excellent agreement with AFM data. Unfortunately, SEM does not allow resolving chains down to the single-molecule level.

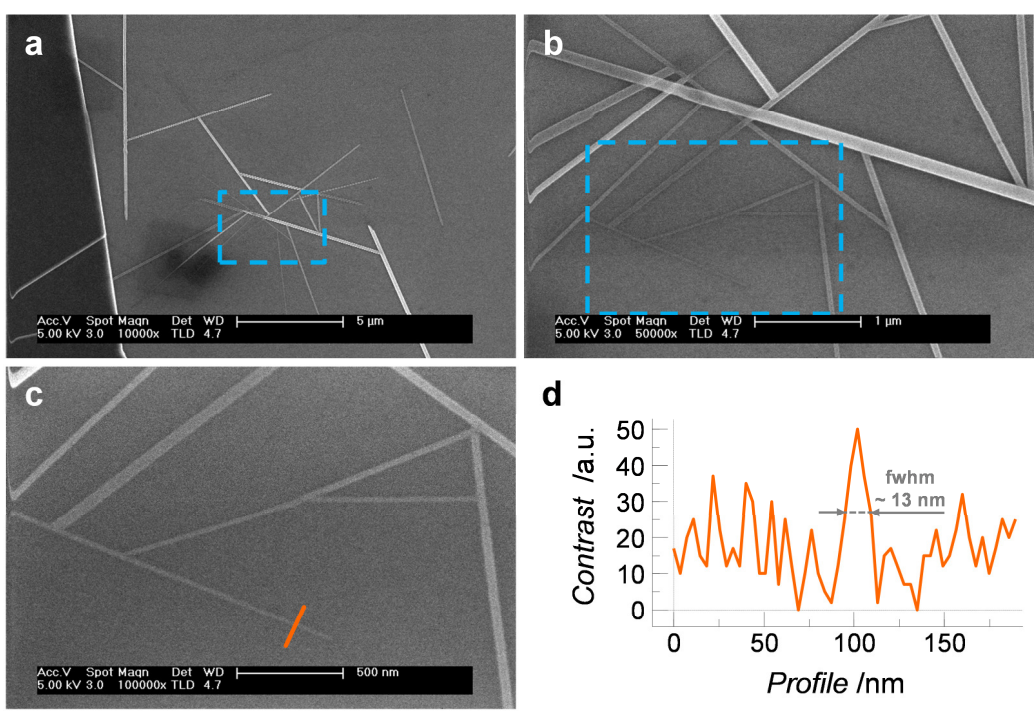

**Figure S1.** (a,c) SEM images with increasing magnifications showing MMX fibres of different sizes. (d) Estimation of the width of the thinnest fibre that we were able to resolve, using SEM, corresponding to the line profile in (c).

**GEOMETRICAL CONSIDERATIONS**

One intrinsic artifact to the AFM technique is the well-known tip dilation. It consists in an apparent widening of thin nanoobjects due to the finite size of the scanning tip. The effects of the finite tip size in AFM topography have been extensively studied (see for instance [1,2]). **Figure S2** shows a geometrical characterization of single chains protruding from a gold electrode. Fig. S2a is the raw topographic AFM image. Notice that the difference in heights between the electrode and the chains hinders their proper visualization. Fig. S2b is the height profile along the solid green line in Fig. S2a. Two individual chains can be discerned, with apparent widths of 15 and 16 nm. The apparent width $w$ of a cylindrical fibre of height $h$ resting on a perfectly flat substrate when imaged with a tip of radius $R$ can be estimated using the expression $w^2 = 8hR$.[3] Hence, considering single chains with height $h \sim 1.2$ nm and a tip radius $R \sim 50$ nm, we can estimate the apparent width of a single chain (the nominal tip radius was 25 nm but, after I-V acquisition, metal coated tips usually become blunt, so that 50 nm is a quite normal value in conductive AFM topographies). We obtain that the apparent width of a single chain is $w \sim 22$ nm, compatible with the experimental chain widths.

It can be observed as well that the mother fibre where the single chains are coming from the gold electrode comprises several chains. Fig. S2c is the height profile along the dashed black line in Fig. S2a. An increase of ~ 12 nm in the gold electrode height indicates that the height of the fibre underneath has to be on that range. This singularity allowed a good electrical contact with the single molecules.

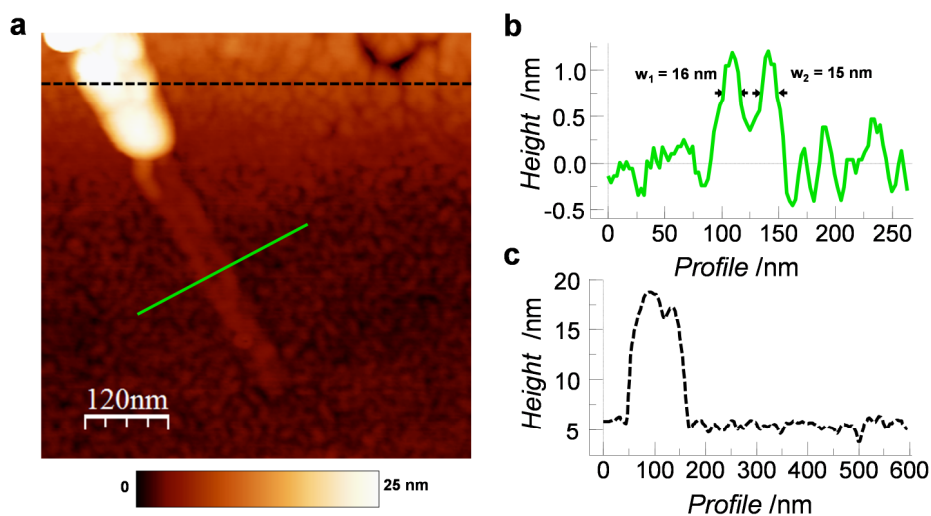

**Figure S2.** (a) Topographic AFM raw image of two single $[Pt_2(EtCS_2)_4I]_n$ chains protruding from a gold electrode. (b) Height profile along the solid green line in (a). Full widths at half maxima are shown. (c) Height profile along the dashed black line in (a).

In order to further illustrate the effects of the finite tip size in MMX chains and to better understand the experimental AFM topographical images in a more quantitative way, we have simulated an image mimicking Figure S2. To this end, first we have simulated the random roughness characteristic of the $SiO_2$ surface (Figure S3a); two single chains with a geometry set by diffraction data, at a distance obtained from Figure S2, taking into account that dilation preserves the distance between chains maxima (Figure S3b); and the gold electrode including the mother fibre where the single chains are coming from (Figure S3c). Figure S3d is the result of adding these images, obtaining an image mimicking the experimental one in Figure S2 but without the tip dilation effects. Figure S3f is the profile along the line shown in 3d. Finally, we have applied a dilation algorithm (Figures S3e and g),[4] which uses a parabolic tip of radius $r$, with $z = (x^2+y^2)/2r$, where $xyz$ are the lateral and vertical coordinates of the image. Dilation treated both tip and chains as hard undeformable bodies. The resulting profile in Figure S3g closely resembles Figure S2b for a tip radius of 50 nm and provides apparent widths compatible with that of single chains observed experimentally.

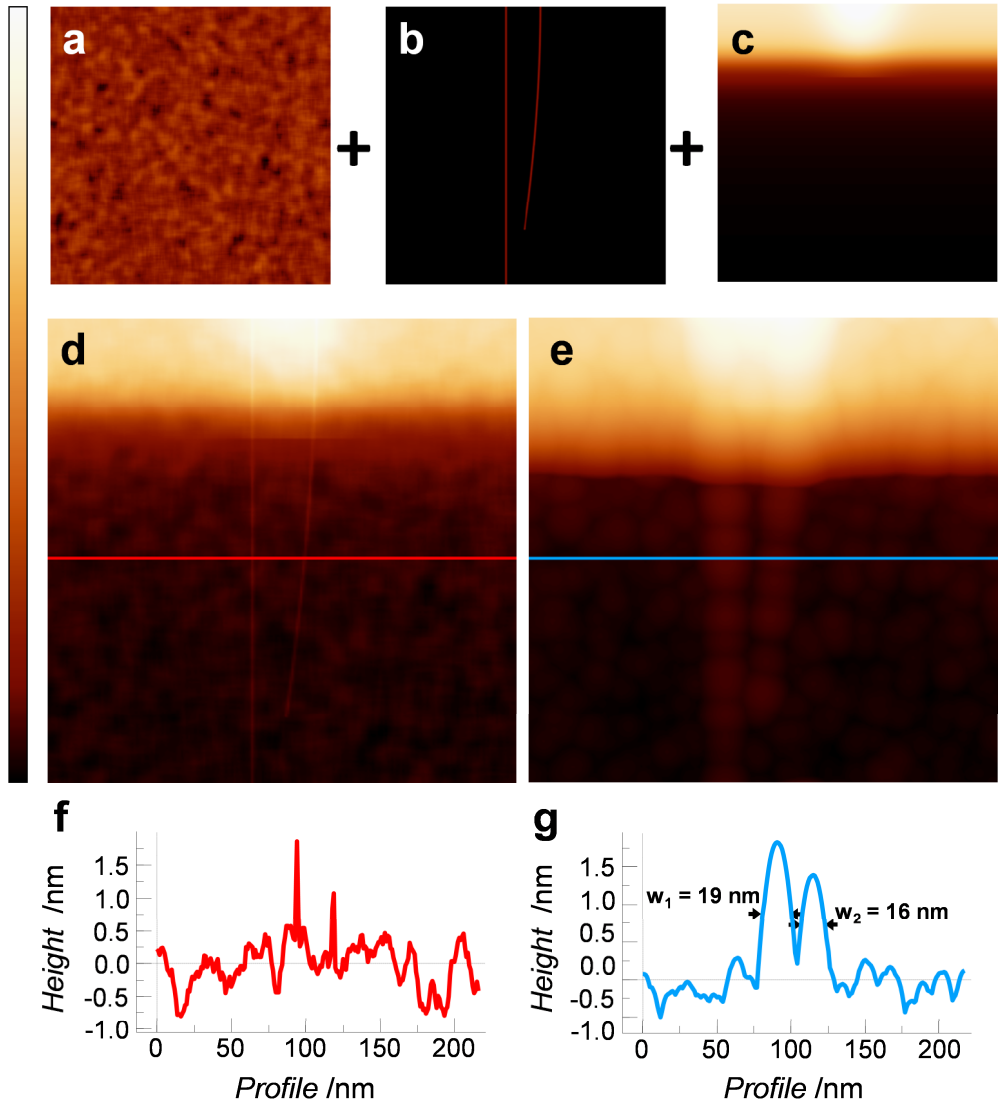

**Figure S3.** Simulated 240 × 240 nm² topography images. (a) SiO$_2$ substrate mimicking its surface roughness by a random noise. (b) Two polymer chains consisting of cylinders with the height expected from diffraction data. The distance between chains was set from Figure S2. (c) Gold electrode (upper part) with the mother fibre where the single chains are coming from. (d) Image mimicking Figure S2 without dilation effects resulting from adding images in (a), (b) and (c). (e) Dilation image obtained from (d) assuming a finite tip radius of 50 nm. (f, g) Profiles from (d) and (e) respectively. Color scale on the left applies for all images, where the height difference between black and white colors are 7.5 nm for (a, b) and 35 nm for the rest.

**CURRENT AND CURRENT DENSITY**

**Table S1.** Length, current and electrical resistance for the $[Pt_2(EtCS_2)_4I]_n$ chains. Current data are extracted from I-V curves at a fixed bias voltage of 0.5 V (these data are plotted in Figure 3 of the manuscript). The uncertainties shown in the table account for instrumental errors in our measurements. However, we think there are other not controlled error sources that may enlarge the magnitude of uncertainty. Because of this, plots in Figures 3 and S4 do not show error bars. In any case, the typical issues present in C-AFM always yield underestimated current values. Therefore, the figures of the electrical resistance/conductance reported herein are always an upper/lower bound for this magnitude.

| L [nm] (± 5 nm) | I [pA]     | R [GΩ]       |
|-----------------|------------|--------------|
| 80              | 2000 ± 100 | 0.25 ± 0.02  |
| 140             | 750 ± 40   | 0.67 ± 0.04  |
| 210             | 230 ± 10   | 2.2 ± 0.1    |
| 280             | 14 ± 1     | 36 ± 3       |

**Figure S4** shows these data in a logarithmic plot to better see the exponential dependence of the resistance with the length.

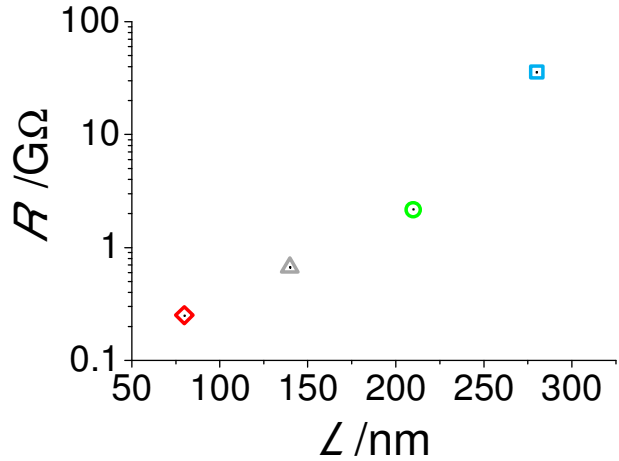

**Figure S4.** Resistance *vs.* length logaritmic plot of data from Figure 3 and Table S1. The exponential dependence of the resistance with the length is clearly visible.

Data in Table S1 were fitted to equation 1 as shown in the main text, $R=R_c+(R_0/N_0)\exp(L/L_0)$, where $R_c$ accounts for the electrode-chain contact resistance, $R_0$ is the quantum resistance ($R_0 \approx 13$ k$\Omega$), $N_0$ is the number of chains contacted by the AFM tip and $L_0$ is the localization length. In our fits, we fixed the term $R_0/N_0$ (considering $N_0 = 2$ in Fig. 3c and $N_0$ from 1 to 4 in Fig. 3d) and we left free $R_c$ and $L_0$. In the case of data from Table S1, this leads to $R_c = 800 \pm 400$ M$\Omega$ and $L_0 = 18.07 \pm 0.02$ nm. All fits in the present work were carried out as explained and yielded $R^2 > 0.99$.

The reproducibility and consistency of conductive AFM data is illustrated in **Figure S5**.

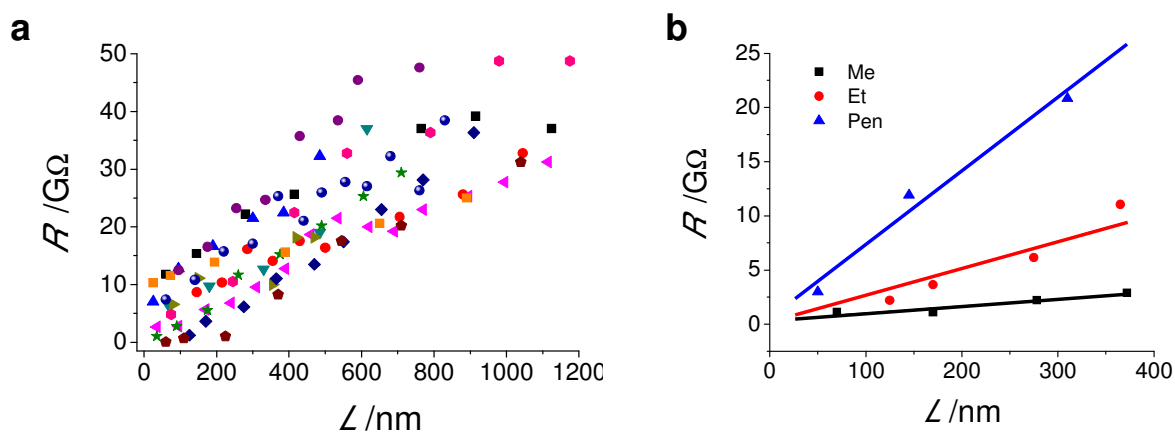

**Figure S5.** (a) Resistance *vs.* Length curves for fibres of $[Pt_2(RCS_2)_4I]_n$ with R = Et diameters ranging from ~ 8 to 14 nm. Each symbol corresponds to a different fibre. Each resistance point was obtained from an I-V curve, at Bias = 1 V. (b) Representative Resistance *vs.* Length curves for fibres of $[Pt_2(RCS_2)_4I]_n$ with R = Me (black), Et (red) and Pen (blue) with diameters ~ 12 nm. Symbols correspond to resistance values obtained from I-V curves, at Bias = 1 V. Lines are linear fits.

From our measurements we can estimate that the current density for a single chain with a diameter of ~ 1.2 nm (Figure 1) and a current of 2 nA measured at 0.5 V, 80 nm away from the gold electrode, is J ~ $10^5$ A $cm^{-2}$. **Figure S6** shows the variation of the current density for the chain shown in Figure 3, as a function of the distance between electrodes. Notice the persistence of significant current even for lengths as long as 280 nm.

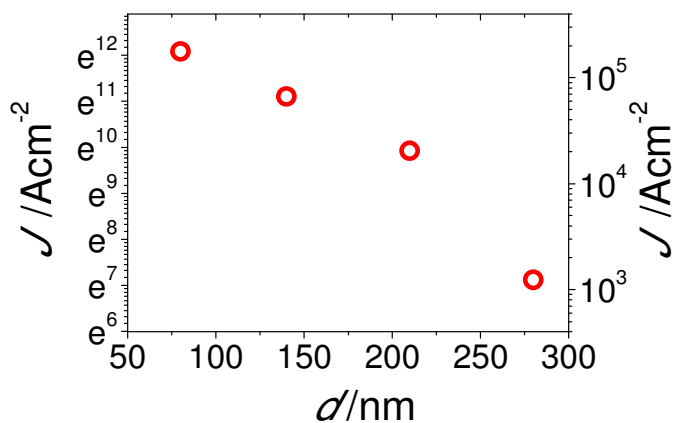

**Figure S6.** Current density at Bias = 0.5 V along a [Pt$_2$(EtCS$_2$)$_4$I]$_n$ single chain as a function of the distance.

We have made Table S2 in order to compare the electrical performance of different molecules published in the literature. As can be seen there, the longest distance reported so far for molecular systems is in the range of ~ 34 nm with experiments that involve ~ 10$^{11}$ molecules, much less stringent than our experiments. This table also puts in perspective the magnitude of MMX conductance found in our work.

**Table S2.** β decay for different molecular systems as reported in the literature.

| Molecular system | $\beta \times 10^{-3}$ [Å$^{-1}$] | # of molecules | Technique | L [nm] | Ref. |
|---|---|---|---|---|---|
| Conjugated oligophenyleneimine structures | 90 – 300 | ~ $10^2$ | AFM | 5 - 8 | [5] |
| DNA | 50 | ~ $10^{11}$ | Electrochemistry | 34 | [6] |
| Co/Fe terpyridine-based molecular wires | 1 - 3 | ~ $10^{11}$ | Hg-drop | 20 | [7] |
| Pt-based MMX polymer chains | 5 | 1-2 | AFM | 280 | This work |